%% file: main.tex
\documentclass[twoside,twocolumn,9pt]{article}
\usepackage{extsizes}
\usepackage[super,sort&compress,comma]{natbib} 
\usepackage[version=3]{mhchem}
\usepackage[left=1.5cm, right=1.5cm, top=1.785cm, bottom=2.0cm]{geometry}
\usepackage{balance}
\usepackage{mathptmx}
\usepackage{sectsty}
\usepackage{graphicx} 
\usepackage{lastpage}
\usepackage[format=plain,justification=justified,singlelinecheck=false,font={stretch=1.125,small,sf},labelfont=bf,labelsep=space]{caption}
\usepackage{float}
\usepackage{fancyhdr}
\usepackage{fnpos}
\usepackage[english]{babel}
\addto{\captionsenglish}{%
  
}
\usepackage{array}
\usepackage{droidsans}
\usepackage{charter}
\usepackage[T1]{fontenc}
\usepackage[usenames,dvipsnames]{xcolor}
\usepackage{setspace}
\usepackage[compact]{titlesec}
\usepackage{hyperref}

\usepackage{verbatim}
\usepackage{ulem}
\usepackage{amssymb}

\newcommand{\tens}[1]{\mathbf{#1}}
\newcommand{\OOmega}{\mathrm{\Omega}}
\newcommand{\PPi}{\mathrm{\Pi}} 

\definecolor{cream}{RGB}{222,217,201}

\begin{document}

\pagestyle{fancy}
\thispagestyle{plain}
\fancypagestyle{plain}{
\renewcommand{\headrulewidth}{0pt}
}

\makeFNbottom
\makeatletter
\renewcommand\LARGE{\@setfontsize\LARGE{15pt}{17}}
\renewcommand\Large{\@setfontsize\Large{12pt}{14}}
\renewcommand\large{\@setfontsize\large{10pt}{12}}
\renewcommand\footnotesize{\@setfontsize\footnotesize{7pt}{10}}
\makeatother

\renewcommand{\thefootnote}{\fnsymbol{footnote}}
\renewcommand\footnoterule{\vspace*{1pt}%
\color{cream}\hrule width 3.5in height 0.4pt \color{black}\vspace*{5pt}} 
\setcounter{secnumdepth}{5}

\makeatletter 
\renewcommand\@biblabel[1]{#1}            
\renewcommand\@makefntext[1]%
{\noindent\makebox[0pt][r]{\@thefnmark\,}#1}
\makeatother 
\renewcommand{\figurename}{\small{Fig.}~}
\sectionfont{\sffamily\Large}
\subsectionfont{\normalsize}
\subsubsectionfont{\bf}
\setstretch{1.125} 
\setlength{\skip\footins}{0.8cm}
\setlength{\footnotesep}{0.25cm}
\setlength{\jot}{10pt}
\titlespacing*{\section}{0pt}{4pt}{4pt}
\titlespacing*{\subsection}{0pt}{15pt}{1pt}

\fancyfoot{}
\fancyfoot[LO,RE]{\vspace{-7.1pt}\includegraphics[height=9pt]{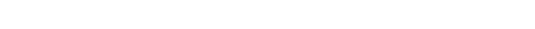}}
\fancyfoot[CO]{\vspace{-7.1pt}\hspace{13.2cm}\includegraphics{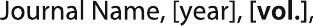}}
\fancyfoot[CE]{\vspace{-7.2pt}\hspace{-14.2cm}\includegraphics{head_foot/RF}}
\fancyfoot[RO]{\footnotesize{\sffamily{1--\pageref{LastPage} ~\textbar  \hspace{2pt}\thepage}}}
\fancyfoot[LE]{\footnotesize{\sffamily{\thepage~\textbar\hspace{3.45cm} 1--\pageref{LastPage}}}}
\fancyhead{}
\renewcommand{\headrulewidth}{0pt} 
\renewcommand{\footrulewidth}{0pt}
\setlength{\arrayrulewidth}{1pt}
\setlength{\columnsep}{6.5mm}
\setlength\bibsep{1pt}

\makeatletter 
\newlength{\figrulesep} 
\setlength{\figrulesep}{0.5\textfloatsep} 

\newcommand{\topfigrule}{\vspace*{-1pt}%
\noindent{\color{cream}\rule[-\figrulesep]{\columnwidth}{1.5pt}} }

\newcommand{\botfigrule}{\vspace*{-2pt}%
\noindent{\color{cream}\rule[\figrulesep]{\columnwidth}{1.5pt}} }

\newcommand{\dblfigrule}{\vspace*{-1pt}%
\noindent{\color{cream}\rule[-\figrulesep]{\textwidth}{1.5pt}} }

\makeatother

\twocolumn[
  \begin{@twocolumnfalse}
\begin{tabular}{m{4.5cm} p{13.5cm} }

 & \noindent\LARGE{\textbf{Coupling Turing stripes to active flows$^\dag$}} \\
\vspace{0.3cm} & \vspace{0.3cm} \\

 & \noindent\large{Saraswat Bhattacharyya,\textit{$^{a}$}  and Julia M. Yeomans\textit{$^{a}$}} \\

& \noindent\normalsize{
We numerically solve the active nematohydrodynamic equations of motion, coupled to a Turing reaction-diffusion model, to study the effect of active nematic flow on the stripe patterns resulting from a Turing instability.
If the activity is uniform across the system, the Turing patterns dissociate when the flux from active advection balances that from the reaction-diffusion process. If the activity is coupled to the concentration of Turing morphogens, and neighbouring stripes have equal and opposite activity, the system self organises into a pattern of shearing flows, with stripes tending to fracture and slip sideways to join their neighbours.
We discuss the role of active instabilities in controlling the crossover between these limits, Our results are of relevance to mechanochemical coupling in biological systems.}


\end{tabular}

 \end{@twocolumnfalse} \vspace{0.6cm}

  ]

\renewcommand*\rmdefault{bch}\normalfont\upshape
\rmfamily
\section*{}
\vspace{-1cm}


\footnotetext{\textit{$^{a}$~The  Rudolf  Peierls  Centre  for  Theoretical  Physics, Department of Physics, University  of  Oxford,  Parks  Road, Oxford OX1 3PU, United Kingdom. E-mail: saraswat.bhattacharyya@physics.ox.ac.uk}}

\footnotetext{\dag~Electronic Supplementary Information (ESI) available: Videos of advected Turing patterns in uniform and extensile-contractile activities, three regimes for each. See DOI: 10.1039/cXsm00000x/}




\section{Introduction}

The term active matter describes systems whose constitutive particles take energy from their surroundings and use this to do mechanical work. Recently, active materials have generated considerable interest, both as examples of systems that remain out of thermodynamic equilibrium and also as a way to understand movement and shape changes in living entities. For example, theories of  active nematics \cite{Doostmohammadi2018,MarchettiReview} have been successful in modelling active turbulence in biological systems \cite{Duclos2017}, identifying topological defects in epithelial cell layers \cite{Saw2017} and growing hydra \cite{Maroudas-Sacks2021}, and explaining spontaneous cellular flow in confinement \cite{Duclos2018}.

Chemical signalling is known to be another important driver for patterning and morphogenetic changes in tissues. Following Turing's seminal paper \cite{Turing1990} showing that coupled reaction-diffusion equations can lead to pattern formation, there have been many interesting results extending his ideas to more closely model biological processes \cite{RefList13, RefList16,Gierer1972,Kondo1616,Howard2011, LANDGE20202}.  
The effect of bio-chemo-mechanical feedback on active shells has been studied recently \cite{RefList1}, as has the coupling of a biphasic model of tissue mechanics with morphogen reaction and diffusion \cite{RefList7}.  Robustness of Turing-pattern formation in biological systems under the effects of noise and feedback is also an active area of research \cite{OxMathBio1}. 


It is now timely to investigate the interplay between flows arising from active stresses and the pattern-formation driven by chemical reaction-diffusion processes. Siero {\it et al.} have discussed  how Turing patterns respond to advective flows \cite{TurAdv1}.
There have been studies on coupling pattern formation to active flow in one dimension \cite{Julicher1}, where the existence of a region of non-trivial pattern formation as an interplay of activity and reaction-diffusion was discovered. This work was extended to show the existence of pulsatory patterns when an activator-inhibitor system of two chemicals was used in higher dimensions \cite{Julicher2}. 

 

In this paper we study the coupling between Turing patterns and active turbulence, the chaotic flow state characteristic of active nematics \cite{Doostmohammadi2018}.
Working in two dimensions, we consider a two-species, reaction-diffusion model which orders into Turing patterns, where each of the chemical species can induce local active stresses. The hydrodynamic equations of motion that describe this system, and our approach to solving them, are introduced in Sec.~2. In Sec.~3, we discuss how activity can affect the Turing stripe patterns formed by the reaction-diffusion system. We find that stripes bend, break up into patches and then completely disappear on progressively increasing the activity. Then, in Sec.~4, we study a specific coupling, where the two chemical species induce activities which are equal in magnitude, but take different signs. This leads to a state with alternating shear flows across Turing stripes, which undergo slipping at longer timescales. In Sec.~5, we discuss the solution for more general activity assignments, highlighting the role of active instabilities.

\section{Hydrodynamic equations of motion}
\label{eqofmotion}

We consider a two-dimensional active nematic, described by the continuum active nematohydrodynamic equations of motion \cite{Doostmohammadi2018, ActNem1, Prost2015}, coupled to reaction-diffusion equations that give rise to Turing patterns \cite{Turing1990}. The coupling is introduced by allowing the strength of the activity to depend on the local concentration of the reactants \cite{Banerjee2019}.

The nematic dynamics can be described in terms of an order parameter $\tens{Q}$, which captures the orientation and strength of the alignment, and a velocity field $\tens{u}$ \cite{degennes_book,Beris1994} . The order parameter is a traceless tensor, defined by 
\begin{equation}
    \tens{Q} =  S_{nem} (2\tens{ nn - I})
\end{equation}
where $S_{nem}$ is the magnitude of the nematic order, and $\tens n$ describes the director alignment. $\tens Q$ evolves according to the Beris-Edwards equation:
\begin{equation}
    \partial_t \tens{Q} + \tens{u}.\nabla\tens{Q} - \tens{S} = \Gamma \tens{H}.
    \label{eqn:Q_Field}
\end{equation}
In Eq.~(\ref{eqn:Q_Field}), $\tens{S}$ is the co-rotation term that accounts for how the elongated particles respond to gradients in the flow, given by
\begin{equation}
    \begin{aligned}
    \tens{S} = (\lambda \tens{E} + \tens{\OOmega}).&\bigg(\tens{Q}+\frac{\tens{I}}{2} \bigg ) + 
    \bigg(\tens{Q}+\frac{\tens{I}}{2} \bigg ).(\lambda \tens{E} - \tens{\OOmega}) \\ 
    - & 2\lambda \bigg(\tens{Q}+\frac{\tens{I}}{2}\bigg)\bigg(\tens{Q}:\nabla\tens{u}\bigg)
    \end{aligned}
    \label{eqn:Corot_term}
\end{equation}
where $\tens{\OOmega}$ and $\tens{E}$ are the vorticity and rate of strain tensors respectively. $\lambda$ is the tumbling parameter, which describes how the particle alignment couples to the flow. \begin{equation}
    \tens{H} =-\frac{\partial \mathcal{F}}{\partial \tens{Q}} + \frac{\tens{I}}{2} Tr\bigg( \frac{\partial \mathcal{F}}{\partial \tens{Q}} \bigg)
\end{equation}
is the molecular field. $\tens{H}$ ensures that, in the absence of flow, the system relaxes to the minimum of a free energy, which we take to have 
the Landau-de Gennes form \cite{degennes_book}
\begin{equation}
    \mathcal{F} = \frac{A}{2} \tens{Q}^2 + \frac{B}{3} \tens{Q}^3 + \frac{C}{4} \tens{Q}^4 + \frac{K}{2} (\nabla \tens{Q})^2
\end{equation}
where the coefficients of the bulk terms $A, B, C$ are parameters of the material system which control the thermodynamic ordering, whereas the final term represents the elastic energy cost associated with spatial distortions in the order parameter, assuming a single elastic constant. 

The equations for the $\tens{Q}$ tensor are coupled to the evolution of the flow field $\tens{u}$:
\begin{equation}
    \nabla \cdot \tens{u} = 0,
    \label{eqn:flow_field_1}
\end{equation}
\begin{equation}
    \rho (\partial_t \tens{u} + \tens{u}\cdot\nabla\tens{u}) = \nabla \cdot \tens{\PPi}
    \label{eqn:flow_field_2}
\end{equation}
where $\rho$ is the density and the stress tensor $\tens \PPi$ includes viscous, elastic and active contributions,
\begin{align}
    \tens \PPi^{viscous} &= 2\eta \tens{E},  \\
    \tens \PPi^{elastic} &= -P\tens{I} + 2\lambda (\tens{Q} + \tens{I}/2)(\tens{Q}:\tens{H}) -  \lambda \tens{H} \cdot (\tens{Q} + \tens{I}/2)  \\ & - \lambda (\tens{Q} + \tens{I}/2)\cdot \tens{H} - \nabla \tens Q \frac{\partial \mathcal F}{\partial \nabla \tens{Q}} + \tens{Q \cdot H} - \tens{H \cdot Q}, \nonumber \\
    \tens\PPi^{active} &= -\zeta \tens{Q}. 
\end{align}
The $\tens \PPi^{active}$ term is responsible for the active stress \cite{SimhaRamaswamy2002, ActStress1, Thampi2016}, with the activity coefficient $\zeta$ controlling the strength of the activity. $\eta$ is the kinematic viscosity, and $P$ is the pressure.

We now add scalar fields $c_A$ and $c_B$ describing the concentrations of two morphogens, A and B, which reside within the active fluid.  Their dynamics is modelled by
\begin{align}
    \partial_t c_A + \tens \nabla \cdot (c_A \tens u) &= D_A \nabla^2 c_A + R_A (c_A, c_B), 
    \label{eqn:c_eqn_1}\\
    \partial_t c_B + \tens \nabla \cdot (c_B \tens u) &= D_B \nabla^2 c_B + R_B (c_A, c_B) \label{eqn:c_eqn_2}
\end{align}
where the flow field $\tens u$ of the active nematic advects each species and $D_A$ and $D_B$ are diffusion constants.  
$R_A$ and $R_B$ are reaction terms.  We use the Schnakenberg model \cite{Turing1990,SCHNAKENBERG1979389,schnak2}, which is a minimal model to generate Turing patterns. In this case, we choose 
\begin{align}    
    R_A (c_A, c_B) &= \gamma(a - c_A + c_A^2 \, c_B) \label{eqn:R_1},  \\
    R_B (c_A, c_B) &= \gamma(b -  c_A^2 \, c_B) \label{eqn:R_2}
\end{align}
where $a$, $b$ and $\gamma$ are model parameters. In the absence of flow the phase space of Eqns.~(\ref{eqn:c_eqn_1}--\ref{eqn:R_2}) consist only of homogenous, stripe-forming, and spot-forming regions \cite{TurPatternTypes}, with the non-trivial Turing patterns appearing for imbalanced diffusion constants,  $d \equiv D_B/D_A \gtrsim 10$ or $\lesssim 0.1$. The concentration profiles of the two species are out-of-phase. 
The characteristic lengthscale of the Turing spots or stripes follows from linear stability analysis as  \cite{TurLengthscale}
\begin{equation}
    L_{Turing} = 2\pi \sqrt{\frac{D_A}{\gamma}} \bigg[ \frac{(a-b)/(a+b)+(a+b)^2}{d-1} + \frac{d+1}{d-1} \sqrt{\frac{2b(a+b)}{d}} \bigg]^{-\frac{1}{2}}.
    \label{eqn:turing_lengthscale}
\end{equation}
In the simulations, we will change the rate of reaction $\gamma$ to control the contrast and width of the Turing patterns.
For small $\gamma$, the rate of reaction is much lower than the rate of diffusion, and we find wide stripes with a small difference in morphogen concentration between consecutive stripes. For larger values of $\gamma$, a strong reaction rate leads to thinner stripes with a higher variation in morphogen concentration.

We solve the equations of motion using the hybrid lattice Boltzmann method \cite{ConvertCoefficients, code_paper}. This involves solving for the flow field, Eqs.~(\ref{eqn:flow_field_1}, \ref{eqn:flow_field_2}), using a lattice Boltzmann approach, and for the orientational order, Eq.~(\ref{eqn:Q_Field}), and concentration, Eqs.~(\ref{eqn:c_eqn_1}, \ref{eqn:c_eqn_2}),  using a first-order finite difference scheme. 
The simulations were run on a grid size of $L_x = 160$, $L_y = 160$ lattice units in the $x$ and $y$ directions respectively. Periodic boundary conditions were used in both $x$ and $y$. Simulations were typically run from a uniform concentration profile with small added noise, or a random initial configuration. However, we checked that the same results hold for a wide variety of initial conditions of concentration and nematic director field.

Parameters used for the active nematic were  $\rho = 40$, $\Gamma  = 1$, $K = 0.15$, $A = -0.1$, $B = 0$, $C = 0.05$, $\lambda = 0.3$, $\eta = 1$, $P = \rho/3$ and $S_{nem} = 1$, and lattice Boltzmann time step,  $\Delta t_{LB}$ =1. The activity $\zeta$ was varied within the range $[0.01, 0.20]$. 

For the reaction diffusion system, we used $D_A = 1.0, D_B = 20.0, a = 0.14, b = 1.41$ corresponding to the stripe forming regime \cite{TurPatternTypes}, and $\gamma$ within the range $[0.1, 1.0]$ to vary the strength and contrast of the Turing stripes. The time step was $ \Delta t_{LB}/200$. The shorter time step is needed to solve for the concentration profile sufficiently accurately in the regime where
the pattern formation mechanism is roughly comparable in strength to the advection i.e.~the Turing lengthscale  is comparable to the lengthscale of the active nematic instability.
 
\section{Turing patterns advected by active flows}
\label{sec:UncorrelatedActivity}

\begin{figure}[h]
    \centering
    \includegraphics[width=0.5\textwidth]{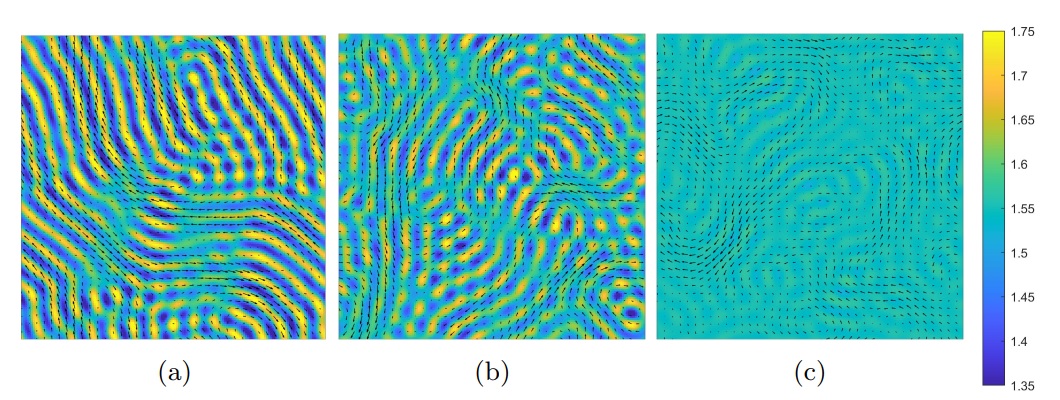}
    \caption{Increasing the strength of activity in a coupled Turing-active nematic system.  
    The colour bar denotes the distribution of the $c_A$ profile and the velocity field is depicted by arrows. (a) $\zeta =0.04$, long bent stripes;  (b)  $\zeta =0.07$, stripe segments; (c) $\zeta =0.17$, no Turing stripes.
  }
    \label{fig:NCD_Stripes_v3}
\end{figure}

For the active reaction-diffusion system
the choice of parameters listed in Sec.~\ref{eqofmotion} creates a Turing pattern of parallel stripes spaced 5.5 lattice units apart with a concentration difference $c_A \in [1.35, 1.75]$, and $c_B \in [0.54, 0.64]$. We first consider the case where the morphogens are advected by the flow field of the active nematic but the nematic is not affected by the presence of the Turing pattern. Stresses due to the  director field of the active nematic create the typical flow vortices of bulk active turbulence \cite{Thampi2016}.

Fig.~\ref{fig:NCD_Stripes_v3} compares examples of the different behaviours of the concentration fields as the activity $\zeta$ is varied.
At low values of activity, Turing stripes emerge and span the entire system. These bend along the direction of the active flow, but are not broken apart (Fig.~\ref{fig:NCD_Stripes_v3}a, Movie 1a). This is possible because 
the vortices generated by the active flux are large in scale, the field lines bend slowly, and the velocities are small. On increasing the activity, the flow is able to overcome the 
ordering and the stripes  break apart to form spots and stripe segments. These patches have a high variation in morphogen concentration (Fig.~\ref{fig:NCD_Stripes_v3}b, Movie 1b) and are advected around by the nematic flow. On further increasing the activity,  Turing stripes no longer form  because the advective flux generated by the activity is sufficient to destroy any patterning generated by the reaction-diffusion flux, henceforth called the {\it Turing flux} (Fig.~\ref{fig:NCD_Stripes_v3}c, Movie 1c). 
The same sequence of phases is seen if the reaction strength $\gamma$ is decreased at fixed activity. Decreasing $\gamma$ decreases the strength of Turing patterns, and consequently lower active flux is required to break them. We have verified that the transition point does not change on changing the elasticity or the flow tumbling parameter of the system.

A P\'eclet number can be defined as the ratio of the timescale of the diffusion-reaction equation (the {\it Turing timescale}) to the advective timescale. The advection is driven by active forces, and the timescale of advection is given by $t_{adv} = {l}/{U} $, where $l$ is a characteristic lengthscale for the active nematic and $U$ is a characteristic velocity \cite{Julicher1}. For an active nematic, $U \sim \zeta l / \eta$ \cite{NemCorrLengths}. Hence, $t_{adv} \sim \eta/\zeta$. When the system is close to forming Turing patterns, the Turing flux is mostly controlled by the reaction term, $\gamma$. Hence, the Turing timescale is proportional to $t_{Turing} \sim (\gamma  )^{-1}$ and  we expect the crossover between different regimes to be controlled by the P\'eclet number, $Pe \sim {\zeta}/{ (\eta \gamma) } $.

\begin{figure}[h]
    \centering
    \includegraphics[width=0.5\textwidth]{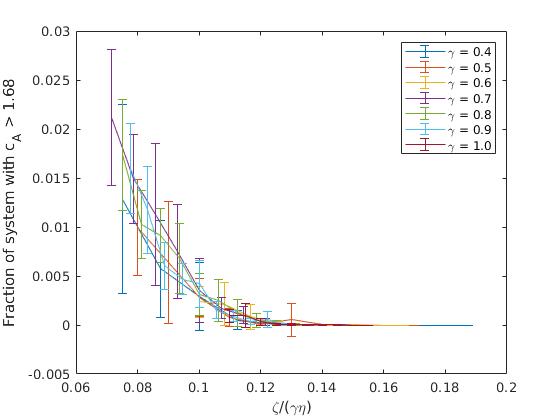}
    \caption{Fraction of active Turing system with concentration of A species above a chosen threshold as a function of P\'eclet number for different values of the Turing reaction strength, $\gamma$. There is a transition at a P\'eclet number $\sim 0.12$ above which clear Turing patterns are destroyed. }
    \label{fig:NCD_Transition_dCA} 
\end{figure}

 To demonstrate this we note that, if the system exhibits Turing patterns, local increases in concentration must be much larger than the fluctuations in concentration.
Therefore the patterns can be characterized by measuring the fraction of the system where the concentration of morphogen A, $c_A$, is greater than a given cutoff value which we take to be 1.68. The maximum concentration is $65\%$ of the Turing value in the absence of advection for this cutoff, however we have checked that the conclusions are independent of the value chosen. Further details can be found in Sec S2 of the Electronic Supplementary Information. This fraction is plotted in Fig.~\ref{fig:NCD_Transition_dCA} as a function of $Pe$ for different values of the Turing reaction strength $\gamma$.  There is a clear crossover to a region where the concentration difference created by the Turing patterns is less than the cutoff, corresponding to the destruction of the stripes, at a P\'eclet number $\sim 0.12$.  Although the error bars are large below the transition due to the noisy nature of an active system, there is a reasonable data collapse with no discernable dependence of the position of the transition on $\gamma$.  

These results are for the Schnakenberg model which gives out-of-phase stripes of the A and B concentrations. A comparison of our results with the in-phase Gierer-Meinhardt stripes \cite{Gierer1972, LANDGE20202}  is discussed in Sec. S3 of the Electronic Supplementary Information.
 
\section{Coupling activity to concentration}

We next investigate the effects of allowing the activity to depend on the local value of the concentrations
$
    \zeta \rightarrow \zeta(c_A, c_B)
$ 
by assigning an activity $\zeta$ if $c_A$ is greater than a threshold value $c_0$ and $\beta \zeta$ if $c_A$  is less than $c_0$. $c_0= 1.55$ per lattice site
is chosen as this corresponds to the mean concentration of the morphogen A. (This is equivalent to choosing an activity $\zeta$ if $c_B$ is less than a threshold value $c_0^B$ and $\beta \zeta$ if $c_B$  is greater than $c_0^B$, by the symmetry of the Turing stripes. $c_0^B= 0.59$ per lattice site
corresponds to the mean concentration of the morphogen B.) We use the same values of the Turing parameters as in Sec.~3, and an activity $\zeta=0.07$, so the unperturbed Turing system forms stripes.

\begin{figure}[h]
    \centering
    \includegraphics[width = 0.5 \textwidth]{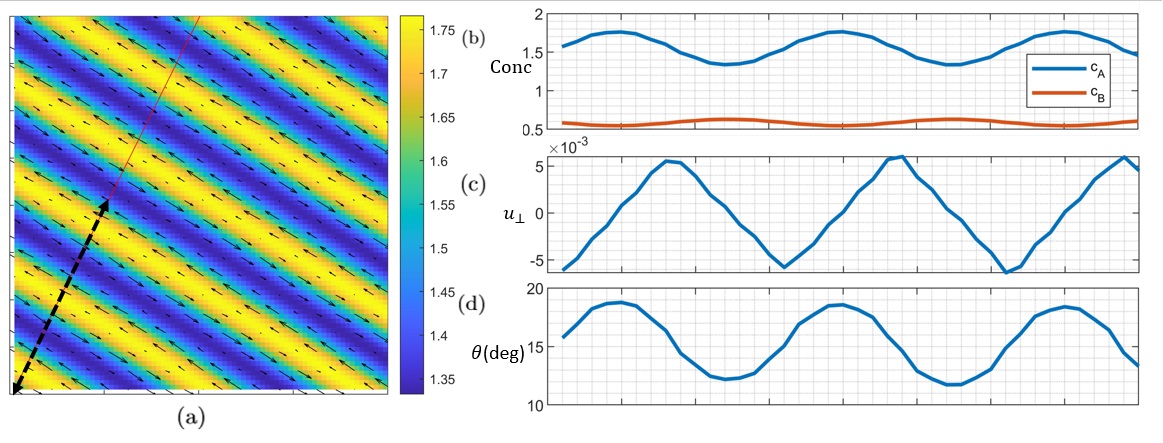}
 \caption{{\bf Alternating extensile and contractile stripes:} (a) Concentration profile of $c_A$ with velocity field marked by arrows. Variation along the black dashed line of (b) concentration fields; 
 (c) velocity  perpendicular to the marked  line. The shear flow has its highest value at the edge of the stripes; (d) angle of the director with respect to the marked line. The angle is maximum in the extensile region.
  }
    \label{fig:EC_Stripes}
\end{figure}

First consider the case  $ \beta = -1$. Regular stripes, with alternating bands of high concentrations of $c_A$ and $c_B$ form at the Turing length scale and span the entire system, as shown in Fig.~\ref{fig:EC_Stripes}a,b. The morphogens respectively generate positive and negative active stresses of equal magnitude. There is Couette flow within each stripe \cite{Couette2020}, with adjacent stripes having opposite shear (Fig.~\ref{fig:EC_Stripes}a,c). 

\begin{figure}[h]
    \centering
    \includegraphics[width=0.5\textwidth]{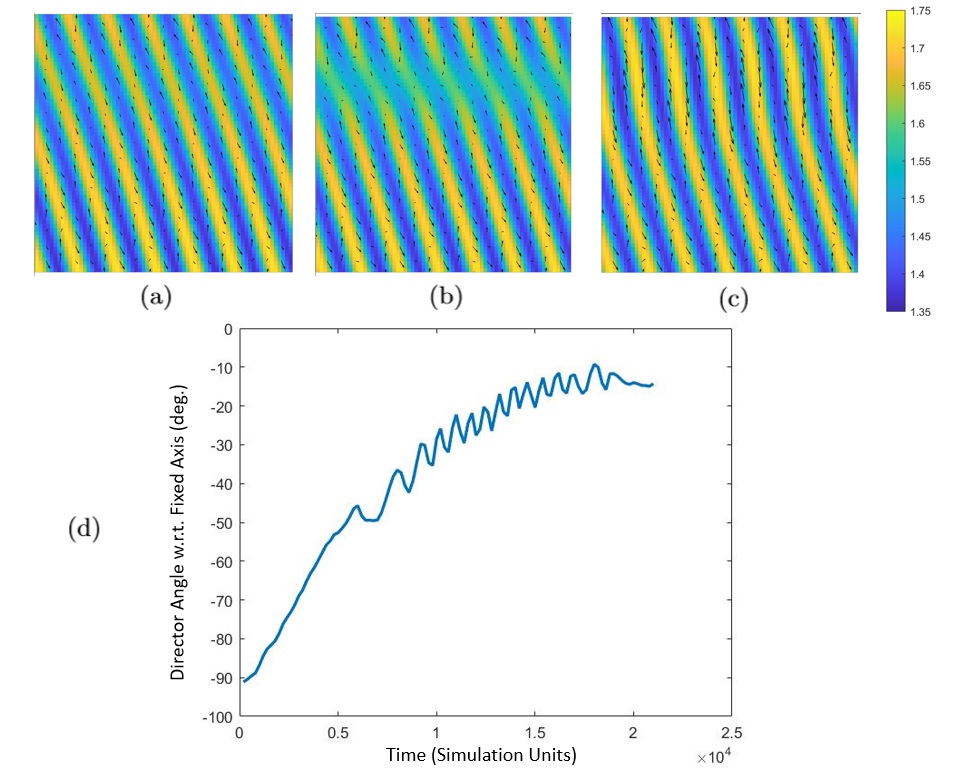}

   \caption{{\bf Slipping of stripes:} 
   (a), (b), (c) sequential time snapshots of a slipping region, taken 200 time steps apart. (d) Variation of the director angle at a point undergoing slipping with time, with respect to a fixed axis. Each slipping event causes the director field to slightly re-set. The superimposed drift at later times occurs because the stripe is slowly rotating.}
    \label{fig:Slipping}
\end{figure}

The angle of the director relative to the normal to the stripe boundaries, $\theta_0$, is approximately constant, but there is a small superimposed sinusoidal variation with a period equal to the width of two stripes (Fig.~\ref{fig:EC_Stripes}d). The director initially orders at an angle $\theta_0 \sim 45^\circ$ to the normal to the stripes. It then slowly rotates towards $\theta_0 \sim 0^\circ$. When the angle reaches approximately $\theta_0 \sim 10^\circ$ to the stripe,
the stripes first bend and then slip along an axis perpendicular to the stripes, at a position determined by the noise. Each stripe breaks at this line and joins with its adjacent neighbour thus increasing the director angle again, and the process repeats  (Fig.~\ref{fig:Slipping}). \\

To understand this behaviour it is helpful to consider first an infinite number of {\bf fixed} stripes of width $L$ with alternating  extensile and contractile activities which we take to lie along the $y$-axis. 
For a director field at an angle $\theta_0$ to the $x$-axis the force acting at the boundaries between stripes due to the change in activity is $ S_{nem }(\nabla \zeta) \sin(2\theta_0)$ parallel to the stripe edges, and $ S_{nem }(\nabla \zeta) \cos(2\theta_0)$ perpendicular to the stripe edges \cite{code_paper}.  The forces alternate in sign on successive stripe boundaries and hence lead to alternating shear flows in neighbouring stripes with proportional shear rates  $\propto S_{nem } (\nabla \zeta)  \sin(2\theta_0)$.

We initiate the simulations with a random noisy director configuration. The director field forms small amplitude sinusoidal oscillations with a wavelength twice the width of a single stripe around a mean angle $\theta_0 \approx 45^{\circ}$. The extensile stripe has a slightly higher director angle than the contractile stripe.

The dynamics of the director as it responds to the shear field is given by \cite{Leslie1987,VanHorn2000,ConvertCoefficients}
\begin{equation}
    \frac{d\theta}{dt} = \frac{u'}{2} (1 + \lambda \cos 2\theta)
    \label{eq:leslie}
\end{equation}
where  $u'$ is the rate of shear across a stripe, and $\theta$ is the angle of the director measured from the normal to the stripe at every point in the system. The angle of the director at the stripe edge is given by $\theta_0$, which sets up the shear flow. Hence, $u' \propto  (\nabla \zeta) \sin(2\theta_0)$.

We are considering the flow-tumbling regime, $0 < \lambda < 1$, and hence the first term in the brackets in Eq.~(\ref{eq:leslie}) dominates. Due to the direction of $ (\nabla \zeta)$, the shear flow tends to turn the director in contractile stripes towards  $\theta=0^\circ$, whereas rotation towards  $\theta = 90^\circ$ is preferred in the extensile stripes.  This is indeed the dynamics observed for very small values of the elasticity $K$. However for larger $K$ this introduces director gradients that are disfavoured by the elastic energy, and instead the whole system rotates slowly towards  $\theta=0^\circ$. 

The reason that  rotation towards $\theta=0^\circ$ is preferred to turning towards $\theta=90^\circ$ follows from noting that there is a small, approximately sinusoidal, variation of the director field superimposed on its constant background value $\theta_0$ (Fig.~\ref{fig:EC_Stripes}d).
This occurs because of the active instability which is then stabilised by the stripe walls  \cite{Joanny2005}.
Therefore the director in the contractile stripes takes a slightly lower average value than that in the extensile stripes. Since $\lambda>0$, it follows from Eq.~(\ref{eq:leslie}) that the rate of rotation is slightly faster 
in the contractile stripes, and  elasticity forces the extensile stripe to follow. As the director rotates, the angle $\theta_0$ becomes lower - this reduces the forces on the boundary ($\sim \sin(2\theta_0)$) and the magnitude of the shear velocity and hence the rotation rate of the director goes down significantly. 
We note that the angle $\theta_0$ remains fixed if the flow aligning parameter is zero, and the nematic aligns parallel to the stripes if it is negative.

We now relax the pinning of the stripes and return to describe the behaviour of striped Turing patterns with alternating extensile and contractile activities.
The reaction-diffusion dynamics promotes each of the Turing stripes to have a fixed width of half the Turing length-scale, given by Eq.~(\ref{eqn:turing_lengthscale}). This width is a function  only of the Turing parameters, and any deviation from it induces a strong reactive flux which drives the system back to the Turing length-scale. 

\begin{figure}[h]
    \centering
    \includegraphics[width = 0.5 \textwidth]{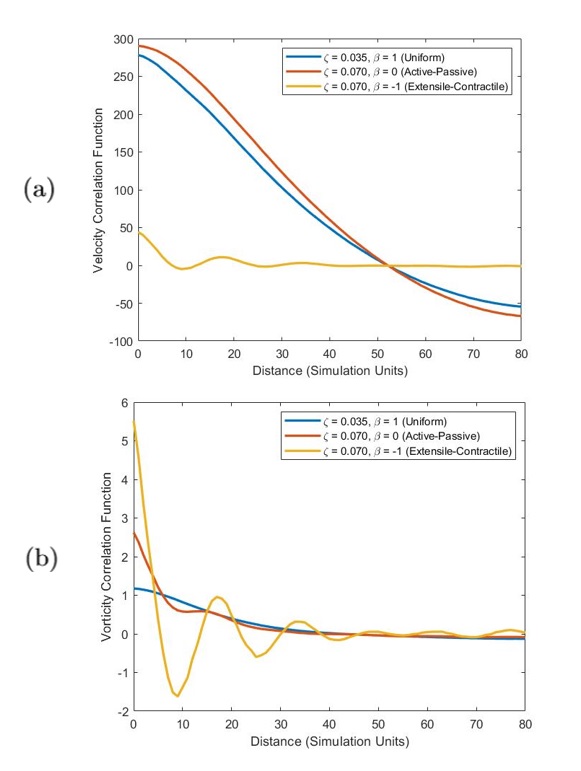}
    \caption{
    Comparison of (a) velocity-velocity and (b) vorticity-vorticity correlation functions for $\beta=1$, $\beta=0$ , and $\beta=-1$. Non-normalized correlation functions are plotted to contrast the magnitudes of the velocity and vorticity.}
    \label{fig:VelVortCorrFunction}
\end{figure}

Initially, Turing patterns form with the stripe boundaries at an angle of approximately $45^\circ$ with respect to the director field.  
At this angle the force at a stripe edge due to the activity gradient is along the stripe edge, and there is no force perpendicular to the edge.
This makes it possible for the Turing system to form stripes without an active force breaking them apart. Just as for the fixed boundary stripes the shear flow results in the director angle decreasing towards $\theta = 0^\circ$. This change in director angle creates forces normal to the stripe boundaries. These forces seek to change the width of the Turing patterns, but are balanced by the reaction-diffusion system, which seeks to oppose this change.

As the director angle decreases the normal forces increase and eventually the stripes start to bend at a random position determined by noise. The bending increases until the stripes collectively slip by fracturing and connecting to the neighbouring stripe (Fig \ref{fig:Slipping}a,b,c, Movie 2a). The reaction-diffusion dynamics then straightens the stripes. The slipping process increases the director angle 
with respect to the new stripes. This re-sets the director angle to a higher value and the process repeats (Fig. \ref{fig:Slipping}d). 

For higher activities, the horizontal forces are strong enough to break up the stripes before slipping can re-set the director angle. However, after the stripes break up, the horizontal forces seeking to break stripes apart disappear, and the stripes can re-form in an uncorrelated direction (Movie 2b), with the director initially making an angle of $45^\circ$ with respect to the normal to the new stripes.
At still higher activities, we observe patches of morphogens A and B advected by the flow, which are unable to form long stripes (Movie 2c). 

We have looked at other possible couplings of $\zeta$ to $c_A$, including linear dependence, sinusoidal dependence, and a case where the Heaviside function is instead smoothed out with a tanh function over 4 lattice sites. In every case, we find qualitatively similar flows to the results reported above.

\section{Active Instability}

We recall that the activity is $\zeta$ if $c_A$ is greater than a threshold value $c_0$ and $\beta \zeta$ if $c_A$  is less than $c_0$, and hence that the Turing system forms stripes with activities $\zeta$ and $\beta \zeta$. In Sec.~3 we considered $\beta=1$ corresponding to an activity independent of concentration, and in Sec~4 we discussed $\beta=-1$ corresponding to alternating extensile and contractile stripes with activities of equal magnitude. We now consider $\beta=0$ which leads to alternating active and passive stripes, and then discuss the crossovers as $\beta$ is varied between these values.

For $\beta=0$ the behaviour with increasing activity follows the pattern observed for $\beta=1$, shown in Fig.~\ref{fig:NCD_Stripes_v3}. However approximately twice the strength of activity is needed to destroy the stripes for $\beta=0$ as active energy is only supplied to half the system.  To demonstrate this we compare the time-averaged velocity-velocity and vorticity-vorticity correlation functions for $\beta=1, \zeta=0.035$,  $\beta=0, \zeta=0.070$, and $\beta=-1, \zeta=0.070$ in Fig.~\ref{fig:VelVortCorrFunction}. 
The long-range nature of the flow field implies that the velocity correlation function is virtually indifferent to the activity distribution for $\beta = 0, 1$, as shown in Fig. \ref{fig:VelVortCorrFunction}a. 
By contrast, the extensile-contractile flow ($\beta = -1$) discussed in Sec. 4 has a different velocity-correlation curve. The magnitude of the velocity is much lower because of the competition between extensile and contractile stresses, and there are oscillations reflecting the underlying stripes. These oscillations appear very clearly in the vorticity-vorticity correlation function (Fig.~5(b)).

We recall that, for $\beta=-1$, the stripes are regular and evolve by slipping along an axis perpendicular to the stripes. Therefore we next allow $\beta$ to take negative values  to understand the crossover from $\beta>0$ where the existence of the stripes is a competition between the active bend instability destroying the stripes and Turing reaction-diffusion re-forming them.

It is helpful to again first consider an infinite number of {\bf fixed} stripes of width $L$ with alternating extensile and contractile activities which we take to lie along the $y$-axis. When the extensile activity is significantly stronger than contractile activity, the system forms bend instabilities across the stripes. On the other hand, when contractile activity dominates, the system forms splay instabilities across the stripes. Director and velocity configurations for instability formation in a typical example are shown in Sec. S4 in the Electronic Supplementary Information. As is well known \cite{Ramaswamy_2007}, extensile instabilities are much easier to form than contractile instabilities, and hence an extensile-dominant system forms instabilities faster than a contractile dominated system. We note that when the strength of extensile and contractile activities is approximately equal, the instability formation is very slow - as each type of instability suppresses the formation of the other.

The situation is similar for Turing stripes. 
However, if extensile and contractile activities are close in magnitude, the angle at the stripe boundary $\theta_0$ can get very low without forming an active instability. If this angle reaches a critical value (approximately $10^\circ$ with respect to the stripe normal), the stripes start collective slipping behavior, connecting to their next neighbours. This re-sets the director angle to a slightly higher value, and the process keeps repeating (Fig. \ref{fig:Slipping}d). Therefore a stronger imbalance in the activity is required to form instabilities across Turing stripes than fixed stripes because of the stabilizing slipping behavior.

 Once a bend or splay instability has formed across one of the stripes, the instability spreads across the entire system and the shear flow across all the stripes gets replaced by bulk active nematic flow patterns of vortices and active turbulence. This breaks the stripes apart into stripe segments and patches. The director and velocity fields set up in this case are similar to the flow of an active nematic with an effective renormalized activity strength. Given a renormalized effective activity strength 
 all the results from Sec. 3 carry forward as $\beta$ increases from -1 once the slipping process (or the balance of activities) fails to overcome the instabilities.

\section{Conclusions}
We have studied the effect of advection by active nematic flows on Turing patterns
finding that active flows destabilize Turing stripes. For very low activity, the Turing stripes bend with the flow lines of the active nematic vortices. For intermediate activities, the stripes break up into spots and line segments which are advected by the active nematic flow. For higher activities, the flow destroys  the Turing patterns. The point at which this occurs is set by a balance between the active advective flux and the Turing flux. 

We then introduced a coupling between the activity of the nematic and the local concentration of the morphogens. In the case where alternating stripes of morphogens have activity of opposite sign (extensile and contractile) but equal strength, the stripes form a series of alternating shear flows across the system, with the highest flow achieved at the boundary between two stripes. 
We observe numerically that the stripes collectively fracture and slip sideways to join their neighbours on long timescales, which stabilizes the stripe system against bend or splay instabilities. 

Phase-separated stripes can also be obtained using mechanisms other than Turing patterns - for instance, similar stripes have been studied in two-dimensional mixtures of an active polar gel and passive isotropic fluid, with an emulsifying surfactant \cite{Bonelli2019}. The flow fields we observe when the activity depends on concentration are reminiscent of the flows observed in polar lamellar systems with contractile activity \cite{Bonelli2019} or external forcing \cite{Negro2019}.

If the activity in all stripes is of the same sign, or if one of the alternating stripes has significantly higher activity than the other, instabilities form across the stripes. These are bend instabilities if extensile activity dominates, and splay instabilities in a predominantly contractile system.  These instabilities destroy the  striped Turing pattern in favour of spots and line segments. The resulting patches of activity set up flows analogous to a bulk active nematic, but with a reduced activity.

Mechanochemical coupling plays a significant role in pattern formation in biological systems \cite{HANNEZO201912, PostTuringBio}. There has also been considerable recent research describing the collective motion of cells in terms of the theories of active nematics \cite{Duclos2017, Yaman2019, Doostmohammadi2018}. Our results contribute to understanding the interplay between activity and pattern formation. In future work it will be interesting to consider more complex reaction-diffusion models, such as those that lead to dynamical patterns \cite{GrayScott, VOLPERT2009267}, or mechanochemical coupling in situations that incorporate shape changes and growth \cite{RefList11}.



\section*{Author Contributions}
SB ran the simulations. Both authors defined the project, analyzed the results and wrote the paper.

\section*{Conflicts of interest}
There are no conflicts to declare.

\section*{Acknowledgements}
We thank Mehrana Raeisian Nejad for useful discussions. SB acknowledges funding from The Rhodes Trust.



\balance


\bibliography{rsc} 
\bibliographystyle{rsc} 

\newpage
\input{main_SM}

\end{document}

%% file: main_SM.tex









\renewcommand{\theequation}{S\arabic{equation}}
\renewcommand{\thefigure}{S\arabic{figure}}
\renewcommand{\thesection}{S\arabic{section}}

\setcounter{equation}{0}
\setcounter{figure}{0}
\setcounter{section}{0}
\onecolumn

\begin{center}
\textbf{\large Electronic Supplementary Information: Coupling Turing stripes to active flows}
\end{center}

\section{Description of the videos}

The seven videos illustrate the results of our simulations. The concentration field $c_A$ is shown as a colour map, with arrows showing the local velocity field. Frames are taken at time intervals of $\Delta t = 200$. The details of the simulation approach, parameters and initial conditions are presented in Sec. 2 of the paper. 

\subsection{Uniform Activity}

A description of the pattern dependence on activity is presented in Sec. 3 of the paper. \\

\textbf{Movie1a}  At lower activities, stripes bend with the flow. $\zeta = 0.04, \gamma = 1$. \\

\textbf{Movie1b}  At intermediate activities, stripes break up into stripe patches and spots.  $\zeta = 0.07, \gamma = 1$. \\

\textbf{Movie1c} At higher activities, stripes dissolve and there are no Turing patterns.  $\zeta = 0.17, \gamma = 1$.\\

\subsection{Extensile-Contractile Activity}

A description of the pattern dependence on activity is presented in Sec. 4 of the paper. \\

\textbf{Movie2a}  At lower activities, stripes collectively slip but do not break apart completely. $\zeta = 0.10, \gamma = 1$. \\

\textbf{Movie2b}  At intermediate activities or lower Turing strengths, stripes break up into stripe patches and spots upon slipping, and then re-form into stripes in an uncorrelated direction. $\zeta = 0.10, \gamma = 0.4$. \\

\textbf{Movie2c}  At higher activities, stripes break up into stripe patches and spots. $\zeta = 0.20, \gamma = 1$. \\

\subsection{Additional movies}

The configurations here are discussed in the Supplementary Materials (Sec S3).\\

\textbf{Movie3a}  Gierer-Meinhardt stripes with alternating extensile and contractile activity. Stripes connect and break at stripe defects, and the highest velocity is at the edge of stripes. $\zeta = 0.07, \gamma = 0.01$.  \\

\section{Characterising the dissolution of Turing stripes}

In the main text, Fig 2, we characterized the crossover from Turing stripes to active turbulence by measuring the fraction of the system where the concentration of morphogen A, cA, is greater than a given cutoff value. We have checked that the data collapses regardless of the choice of cutoff. We have also checked that the data collapses for other cutoff-independent parameters - the amplitude of patterns, and the standard deviation of the concentration profile. We have verified that the transition point does not change on changing the elasticity or the flow tumbling parameter of the system. 

\begin{figure}[htp]
    \centering
    \includegraphics[width = 0.6\textwidth]{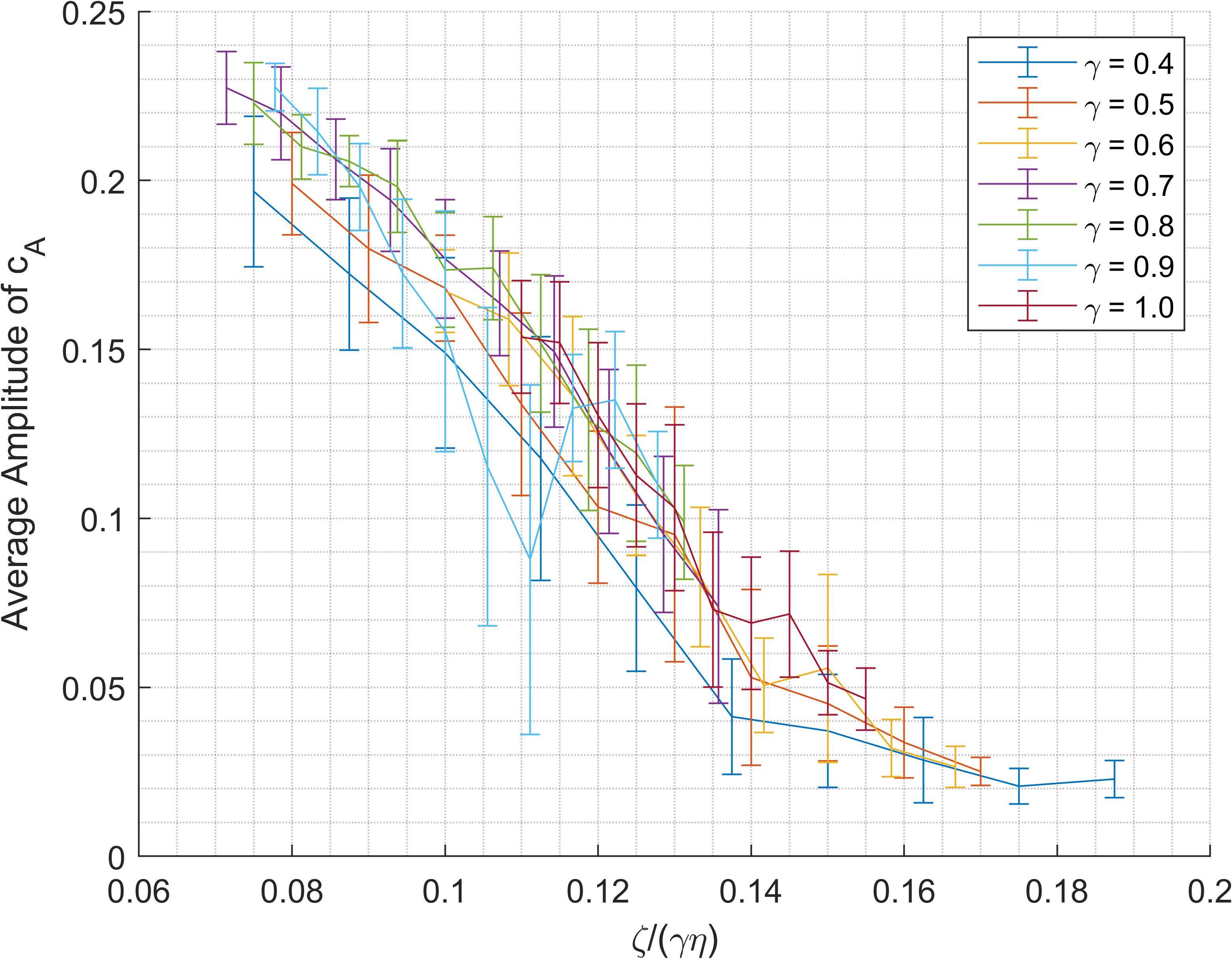}
    \caption{Average amplitude of concentration of species A in an active Turing system as a function of P\'eclet number for different values of the Turing reaction strength, $\gamma$. The amplitude of Turing patterns decreases as a function of P\'eclet number.}
    \label{fig:Sch_amp}
\end{figure}

    Fig.~\ref{fig:Sch_amp} uses a different measure, the average amplitude of the Turing patterns to demonstrate the collapse with P\'eclet number. We define the amplitude of a snapshot of the system as the difference between the highest concentration and the lowest concentration of $c_A$ in that frame. We average over uncorrelated time snapshots separated 200 lattice Boltzmann timesteps apart to calculate the average amplitude. Small fluctuations in the concentration profile appear as noise in the average amplitude. We have found it more insightful to look at only large concentration differences above a cutoff, where we see clear Turing patterns.

\section{Gierer-Meinhardt stripes}

\begin{figure}[htp]
    \centering
    \includegraphics[width = 0.5\textwidth]{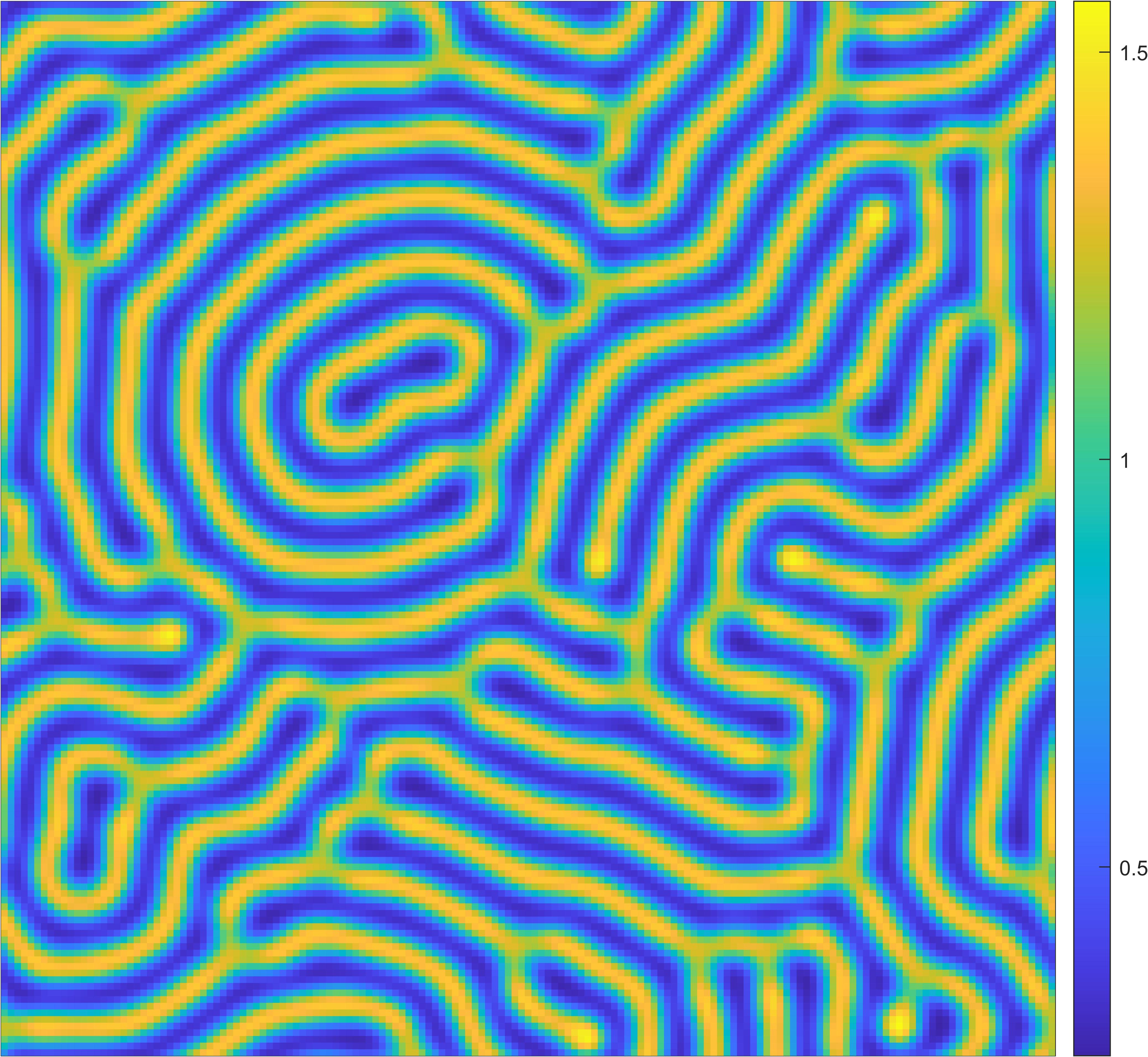}
    \caption{Gierer-Meinhardt static stripe configuration with defects.}
    \label{fig:GM_config}
\end{figure}

We can also use reaction-diffusion systems to generate in-phase Turing patterns. For this, we use the Gierer-Meinhardt model, choosing the reaction terms
\begin{align}    
    R_A (c_A, c_B) &= \gamma \bigg( \big(X_0 + \frac{c_A^2}{1+K_A c_A^2} \big) / c_B - \mu_A \,c_A \bigg) \label{eqn:R_1},  \\
    R_B (c_A, c_B) &= \gamma \bigg( c_A^2 - \mu_B \,c_B \bigg) . \label{eqn:R_2}
\end{align}
We use parameters $\mu_A = \mu_B = 5, \gamma = 0.01, D_A = 0.2, D_B = 10, X_0 = 0.1, K_A = 0.25$. This creates Turing patterns with parallel stripes spaced 20 lattice units apart, with a concentration difference $c_A \in [0.27,1.56]$, and $c_B \in [0.14,0.20]$.  In contrast to the Schnakenberg case, the Gierer-Meinhardt stripes do not  align globally, but have a fixed concentration profile with defects where stripes with different orientations join together (Fig. \ref{fig:GM_config}). 

\begin{figure}[htp]
    \centering
    \includegraphics[width = 0.5\textwidth]{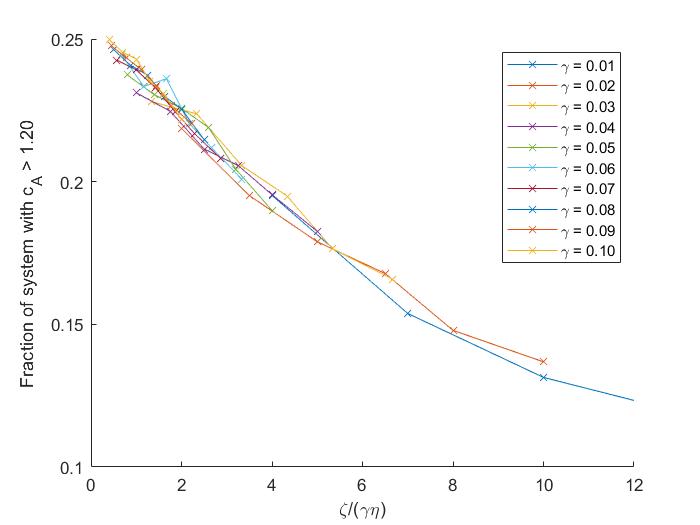}
    \caption{Fraction of active Turing system with concentration of species A above a chosen threshold as a function of P\'eclet number for different values of the Turing reaction strength, $\gamma$ - for Gierer-Meinhardt stripes. The data collapses with P\'eclet number, however the stripes do not dissolve in the range of activity parameter that we were able to simulate. }
    \label{fig:GM_Sec3}
\end{figure}

 We have observed that the average amplitude of the pattern collapses when plotted against P\'eclet number for different values of the Turing reaction strength, $\gamma$ (Fig. \ref{fig:GM_Sec3}), hence the results from Section 3 do not change significantly. However, we find that the Gierer-Meinhardt stripes have a much higher concentration difference than the Schnakenberg stripes, and we do not observe the dissolution of stripes in the range of activity parameter that we are able to simulate. The velocity profile is along the edges of the stripes as described in Section 4. The stripe concentration profile has defects (i.e.~places where stripes join) even in the absence of activity, and the stripes slip at these defects on adding activity (Movie 3a). There is no regular shear state.

\section{Active Instabilities}

\begin{figure}[htp]
    \centering
    \includegraphics[width = 0.6\textwidth]{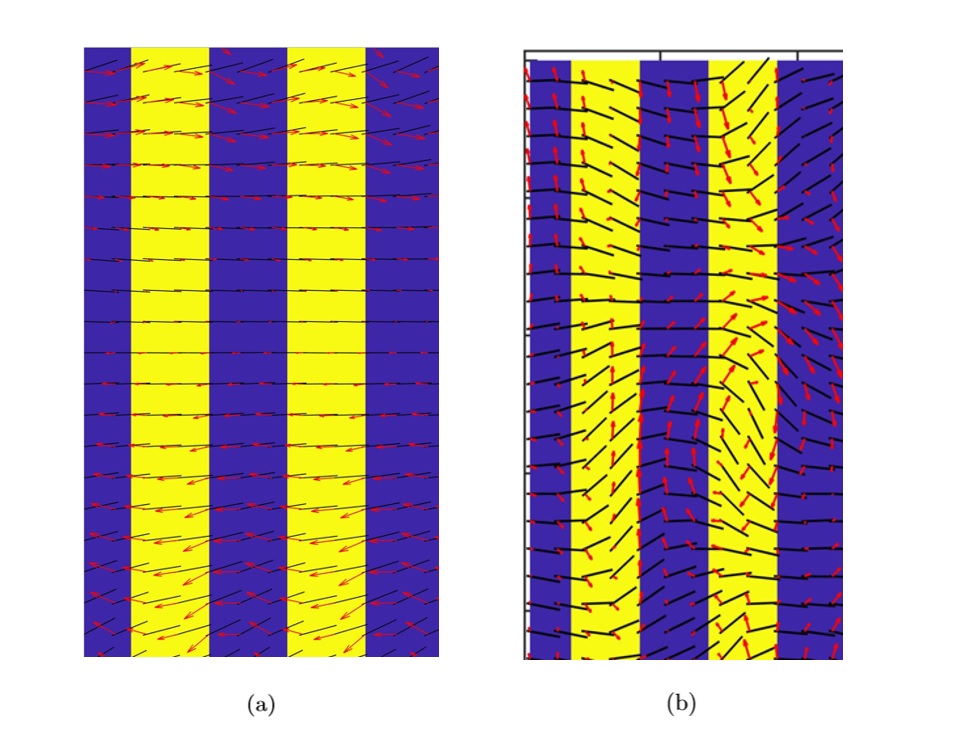}
    \caption{Director (black lines) and velocity (red arrows) configurations for (a) a bend instability: $\zeta_1 = 0.225, \zeta_2 = -0.075$  (b) a splay instability: $\zeta_1 = -0.15, \zeta_2 = 0$.}.
    \label{fig:S5}
\end{figure}

We show the director and velocity fields for bend and splay instabilities in situations where the stripes have different signs and magnitudes of activity coefficient. An example of an extensile-dominant system is shown in Fig. \ref{fig:S5}a , and Fig. \ref{fig:S5}b shows a contractile-dominant system.

